\documentclass[fleqn,10pt]{olplainarticle}

\title{Emergent Democracy}
\date{2004 \\ Version 3.2}
\author{Joichi Ito\thanks{Written with the help of numerous people on the Internet}}
\editor{Jon Lebkowsky}
\keywords{Emergent Democracy, Social Media, Internet}

\begin{abstract}
This essay argues that a new form of democracy --- an ``Emergent Democracy'' --- will develop as a result of the use of Internet communication tools and platforms such as blogs. The essay explores a variety of tools available and explores the history of democracy, modern experiments with democracy and how these tools might support democracy. The essay also explores concerns as these new tools emerge. These issues include concerns such as privacy and the societally negative use of these tools by corporations, totalitarian regimes and terrorists.
\end{abstract}

\begin{document}

\flushbottom
\maketitle
\thispagestyle{empty}

\section*{Preface}

Joichi Ito is an intensively hip fringeophiliac investor based in Tokyo and cyberspace; he is also the nexus of a loose community of social software entrepreneurs and hackers along with tech-focused academics, writers and miscellaneous travelers who hang out online in Joi’s chat room, and who converse asynchronously via weblog posts and comments.

In March 2004 Joi started thinking about the democratic potential of weblogs. He invited anyone reading his weblog to join a ``happening,'' a teleconference augmented by a chat room (for visual feedback) and a wiki (for collaborative note-taking and annotation). The first happening led to a second, after which Joi wrote the Emergent Democracy essay and circulated it as a Word document. Someone else posted the document to a system called Quicktopic that includes a forums-based document review capability. Joi encouraged anyone to review the document and post comments. He incorporated these comments and others that he received on copies of the Word document he’d sent around.

The resulting version was posted on Joi’s wiki, a collaborative workspace where anyone could edit or add text.

In all this the idea was to use an open, democratic process to create a document about social software, especially weblogs, and democracy. There are two ways to look at this. First, as publishing. Weblogs are simple content management systems for publishing to the web, and political web logs could be compared to the tracts published by early activists like Tom Paine. Second, weblogs are conversations. There’s a lot of call and response among bloggers, through their blogs and through embedded systems for posting comments. As platforms for conversation, weblogs can support the kind of discussion and debate that are so crucial to democratic systems.

The essay is the result of an experiment, but it is also a good overview of current thinking about social software and its relevance to political life. It’s even more interesting given the nature of the virtual community that produced it. Joi is international (he was raised and schooled in both the U.S. and Japan), and he spends much of his life online establishing casual friendships with other cyberspace denizens who are spread around Japan, the U.S., and other countries. When Joi thinks of democracy, he is thinking of both the U.S. and Japan, since he has roots in both places.

I was involved in the process of creating the essay, and I have revised the version included here to enhance readability and clarify some assumptions that were left hanging in the original. An earlier version is posted at \url{http://joi.ito.com/static/emergentdemocracy.html}.

- Jon Lebkowsky

\section*{Introduction}

Developers and proponents of the Internet have hoped to evolve the network as a platform for intelligent solutions which can help correct the imbalances and inequalities of the world. Today, however, the Internet is a noisy environment with a great deal of power consolidation instead of the level, balanced democratic Internet many envisioned.

In 1993 Howard Rheingold wrote \cite{rheingold1993virtual},

\begin{quotation}
We temporarily have access to a tool that could bring conviviality and understanding into our lives and might help revitalize the public sphere. The same tool, improperly controlled and wielded, could become an instrument of tyranny. The vision of a citizen-designed, citizen-controlled worldwide communications network is a version of technological utopianism that could be called the vision of ``the electronic agora.'' In the original democracy, Athens, the agora was the marketplace, and more--it was where citizens met to talk, gossip, argue, size each other up, find the weak spots in political ideas by debating about them. But another kind of vision could apply to the use of the Net in the wrong ways, a shadow vision of a less utopian kind of place--the Panopticon.
\end{quotation}

Rheingold has been called naïve \cite{rheingold2000virtual}, but it is clear that the Internet has become a global agora, or gathering place. Effective global conversation and debate is just beginning. We are on the verge of an awakening of the Internet, an awakening that may facilitate the emergence of a new democratic political model (Rheingold's revitalization of the public sphere). However it could also enable the corporations and governments of the world to control, monitor and influence their constituents, leaving the individual at the mercy of and under constant scrutiny by those in power (an electronic, global Panopticon).

We must influence the development and use of these tools and technologies to support democracy, or they will be turned against us by corporations, totalitarian regimes and terrorists. To do so, we must begin to understand the process and implications necessary for an Emergent Democracy. This new political model must support the basic characteristics of democracy and reverse the erosion of democratic principles that has occurred with the concentration of power within corporations and governments. New technologies can enable the emergence of a functional, more direct democratic system which can effectively manage complex issues. Viable technologies for direct democracy will support, change or replace existing representative democracies. By direct democracy, we don’t mean simple majority rule, but a system that evolves away from the broadcast style of managed consensus to a democratic style of collective consensus derived from “many-to-many” conversations.

\section*{Democracy}

The dictionary defines democracy as ``government by the people in which the supreme power is vested in the people and exercised directly by them or by their elected agents under a free electoral system.'' With the Gettysburg address, Abraham Lincoln gave us a more eloquent definition, government ``of the people, by the people, and for the people.''

A functional democracy is governed by the majority while protecting the rights of minorities. To achieve this balance, a democracy relies on a competition of ideas, which, in turn, requires freedom of speech and the ability to criticize those in power without fear of retribution. In an effective representative democracy power must also be distributed to several points of authority to enable checks and balances and reconcile competing interests.

\begin{quotation}
The failure of democracy to scale is also not complicated to understand. The founding fathers of this country, the ``egalitie, fraternitie and libertie'' of France and most other liberals that moved society toward freedom and liberty in the 1700's could not have been expected to visualize the growth of populations, radical evolution of science, vast increases of technology and incredible increases in mobility of information, money, goods, services and people. Nor could they know or visualize the topography of countries such as the United States, Canada and China, or continents such as Africa, Northern Europe, Russia or Latin America. They laid out such vast topography to the best of their ability on grids that bore no resemblance to the reality of the environment or to the huge increases in scale of population commerce and government. In the main, they did not foresee a need for the right to self-organize -- to adjust scale and degrees of separation as such increases occurred \cite{Ito2003} (Dee Hock)
\end{quotation}

As the issues facing government become larger and more complex, new tools are enabling citizens to self-organize more easily. It is possible that such tools will enable democracies to scale and become more adaptable.

A democracy is ideally governed by the majority and protects the rights of the minority. For a democracy to perform this properly it must support a competition of ideas, which requires critical debate, freedom of speech and the ability to criticize power without fear of retribution. In a true representative democracy the power must be distributed into multiple points of authority to enable checks and balances.

\subsection*{Competition of ideas}

Democracy is itself an incomplete and emergent political system, and must, by its nature, adapt to new ideas and evolving social standards. A competition of ideas is essential for a democracy to embrace the diversity of its citizens and protect the rights of the minority, while allowing the consensus of the majority to rule.

This foundation was considered so fundamental to the success of democracy, that the First Amendment to the United States Constitution enumerates three rights specifically to preserve the competition of ideas: the freedoms of speech, of the press, and of peaceable assembly.

\subsection*{Critical debate and freedom of speech}

The competition of ideas requires critical debate that is widely heard. Although we have many tools for managing such debate, increasingly there are barriers to our engaging in it at all.

\subsubsection*{The commons \cite{lessig2002future}}

Effective debate requires a shared set of references and metaphors. The expansion of culture and knowledge depends on linguistic and conceptual shorthand based on shared knowledge and experience. Collaborative, innovative discussion is impossible if every item must be expanded and reduced to so-called first principles. This body of knowledge, experience and ideas has come to be known as a commons.

\begin{quotation}
If nature has made any one thing less susceptible than all others of exclusive property, it is the action of the thinking power called an idea, which an individual may exclusively possess as long as he keeps it to himself; but the moment it is divulged, it forces itself into the possession of every one, and the receiver cannot dispossess himself of it. Its peculiar character, too, is that no one possesses the less, because every other possesses the whole of it. He who receives an idea from me, receives instruction himself without lessening mine; as he who lights his taper at mine, receives light without darkening me.

That ideas should freely spread from one to another over the globe, for the moral and mutual instruction of man, and improvement of his condition, seems to have been peculiarly and benevolently designed by nature, when she made them, like fire, expansible over all space, without lessening their density in any point, and like the air in which we breathe, move, and have our physical being, incapable of confinement or exclusive appropriation. (Thomas Jefferson)
\end{quotation}

Another aspect: the Internet may be considered a commons or public network, though there is persistent threat of enclosure (transferring resources from the commons to individual ownership) based on enforcement of intellectual property and distribution rights. However no one owns the Internet, and no single national entity has jurisdiction, so it remains an open, accessible platform for all kinds of activity, including the evolution of the social commons described above.

\subsection*{Emerging Limits on Debate}

The competition of ideas requires critical debate that is widely heard, and open to a diverse set of participants. Although we have many tools for conducting such debate, increasingly there are barriers to our engaging in it at all.

Even though ideas are not, in theory, subject to copyright, trademark or patent protection, increasingly draconian intellectual property legislation practically limits the scope and meaning of fair use and the flow of innovation, thereby having the same effect as if ideas were property owned and controlled by corporations. This includes the code inside computers and networks, which controls the transmission or reproduction of information. It includes spectrum allocation, determining whether it is shared by individuals or allocated to large corporations broadcasting protected intellectual property \cite{IPMeme}. The effect of these laws is broad, especially given the chilling effect in the fear of lawsuits.

As the notion of intellectual property continues to grow in scope, more and more of what was one part of common knowledge is becoming the property of corporations. As the infrastructure for communication becomes more tuned to the protection of property than the free spread of ideas, the capacity for critical debate is severely constrained.

\subsection*{The Role of Media}

The competition of ideas has evolved as technology has advanced. For example, the printing press made it possible to provide more information to the masses and eventually provided the people a voice through journalism and the press. Arguably, this has been replaced by the voice of mass media operated by large corporations. As a result, there is less diversity and more internalization of the competition of ideas.

Weblogs are web sites that include links and personal commentary published in reverse chronological order. Often alled “blogs” for short, weblogs have become a standard for online micropublishing and communication, thanks to the development of several simple content management systems that support the weblog format \cite{01}. In The Weblog Handbook, Rebecca Blood catalogs several types of weblogs, noting that the classic type is the filter, a type of weblog that filters web content by some criterion (often the weblog author’s interest), essentially a collection of links that point to web pages and web sites, and a usually brief description of the link and why it is interesting. Weblogs and other forms of filtering, coupled with many of the capture and transmission technologies discussed by Steve Mann, author of “Wearable Computing: Toward Humanistic Intelligence” \cite{02} may provide a better method of capturing and filtering relevant information. At the same time, they may suppress information where the privacy damage exceeds the value to the public.

An example of weblogs exceeding the ability of the mass media to identify relevant information is the case of Trent Lott. The national media covered briefly his racist comments during Strom Thurmond's 100th birthday party. After the national media lost interest, the weblogs continued to find and publicize evidence of Lott's hateful past until the mass media once again took notice and covered the issue in more depth \cite{03}.

The balance between what’s relevant and what’s not relevant is culturally biased and difficult to sustain. We need mechanisms to check filters for corruption and weighted perspectives. A variety of checks and balances and a diversity of methods and media can provide the perspectives we need for a balanced view of current events.

Blogs may be evolving to become more than filters; they may be replacing traditional news sources, according to journalist Jay Rosen:

\begin{quotation}
``Blogs are undoing the system for generating authority and therefore credibility of news providers that's been accumulating for well over 100 years. And the reason is that the mass audience is slowly, slowly disappearing. And the one-to-many broadcasting model of communications--where I have the news and I send it out to everybody out there who's just waiting to get it--doesn't describe the world anymore. And so people who have a better description of the world are picking up the tools of journalism and doing it. It's small. Its significance is not clear. But it's a potentially transforming development... I like [it] when things get shaken up, and when people don't know what journalism is and they have to rediscover it. So in that sense I'm very optimistic.'' \cite{04}
\end{quotation}

\subsection*{Privacy}

Whether a system is democratic or otherwise, people or groups with power or wealth often see no benefit in keeping the general population well informed, truly educated, their privacy ensured or their discourse uninhibited. Those are the very things that power and wealth fear most. Old forms of government have every reason to operate in secret, while denying just that privilege to subjects. The people are minutely scrutinized while the powerful are exempt from scrutiny \cite{05}. (Dee Hock) We can’t expect support where power and wealth are concentrated, beyond lip service, for greater, truly meaningful citizen participation in governance. Greater democracy requires that we work constantly to build and sustain structures for effective democratic participation.

In addition to the technical and legal ability to speak and engage in critical debate, citizens must be allowed to exercise this ability without fear of retribution from government or from other institutions where power and wealth are concentrated (e.g. corporate entities). In the increasingly sophisticated world of massive databases and systematic profiling of individuals, the protection of those citizens willing to question and challenge power must be assured. In a networked society where each individual’s data has value, we should consider a definition of rights by which each individual owns and manages data relevant to his identity.

It is essential to understand the difference between personal privacy and transparency. While individuals have a right to privacy, powerful institutions must operate transparently, so that abuses of power are not concealed by veils of secrecy.

In one of the earliest critiques of a national ID card proposal (January 1986), Professor Geoffrey de Q Walker, now dean of law at Queensland University, observed:

\begin{quotation}
One of the fundamental contrasts between free democratic societies and totalitarian systems is that the totalitarian government [or other totalitarian organization] relies on secrecy for the regime but high surveillance and disclosure for all other groups, whereas in the civic culture of liberal democracy, the position is approximately the reverse \cite{06}.
\end{quotation}

Steve Mann presents the notion of sousveillance \cite{07} as a method for the public to monitor the established centers of power and provide a new level of transparency. Traditionally, this has been the role of the press, but the press is decreasingly critical and vigilant, instead focusing on sensational stories, propaganda, and “infotainment.”

\subsection*{Direct Democracy and Scale}

The concept of direct democracy, where citizens are directly responsible for their own governance - originated in Athens, Greece, around the fifth century. Though Athenian democratic governance was direct, it was also limited. Only males born of an Athenian mother and father participated. According to Professor Paul Cartledge, ``the citizen body was a closed political elite'' \cite{08} Of a population of 250,000, an average of 30,000 were eligible to participate, i.e. a mere 12\%, a relatively small group with little diversity.

Common supposition is that direct democracy is not feasible for large, diverse populations. There are practical and technical issues: how do you coordinate ongoing large-scale decision-making that is effective for a broad populace? How can the general population digest and comprehend the complexities involved in running a large state requiring deep understanding of the issues, specialization, and a division of labor. Representative democracy, wherein elected representatives of the people are chosen through a voting mechanism, is considered by most to be the only possible way to manage a democracy of significant scale.

Democratic nations generally adopt republican form of representative democracy, formed in reaction to governments where leadership was hereditary (monarchy). The hereditary model was abandoned and leaders were periodically appointed under a constitution. Republics now tend to be representative democracies, where leaders are periodically elected by citizens, and all adults with few exceptions have an opportunity to vote. Representative democracy allows leaders to specialize and focus on the complex issues of governance, which an uneducated and uninterested general population could not be expected to grasp. Representative democracy is considered more practical than direct democracy, which becomes increasingly difficult with larger and more diverse populations.

The failure of democracy to scale is easy to understand. The founding fathers of the United States, the ``égalité, fraternité and liberté'' of France, and most other liberals who moved society toward freedom and liberty in the 1700's, could not foresee the accelerated population growth over the following two centuries. The couldn’t predict the radical evolution of science, the rapid development of technology and the pronounced increases in mobility of information, money, goods, services and people. Nor could they know or visualize the topography of countries such as the United States, Canada, and China, or continents such as Africa, Northern Europe, Russia, or Latin America. Evolving nations were laid out on maps that bore little resemblance to the reality of the environment, and were not predictive of the huge increases in scale of population, commerce, and government. In the main, no one foresaw a need for the right to self-organize -- to adjust scale and degrees of separation as such increases occurred \cite{Ito2003}.

As the issues facing government have become more complex, social technologies have emerged that enable citizens to self-organize more easily. These technologies may eventually enable democracies to scale and become more adaptable and direct.

As the voting mechanism becomes more organized and the difficulty of participating in critical debate increases, forms of influence are increasingly relevant and detrimental to the balance of power. Elected representatives attend more readily to those who have the power to influence the voting mechanism and the public debate; these are often minorities who have more financial influence or the ability to mobilize large numbers of motivated people through ideological or religious channels. Extremists and corporate interests can become dominant, and a “silent majority” may have little input into the selection of representatives or the critical debate \cite{09}.

A variety of groups have been successful in polling the silent majority and amplifying its opinions to provide support for moderate politicians on policy issues. One such group , One Voice \cite{10}, conducts telephone and Internet polls of average Israeli and Palestinians, most of whom are in favor of peace.. The organization amplifies their opinions by publishing poll results in reports and the mass media. This method of bypassing the traditional methods of influencing representatives is a form of direct democracy, which is becoming increasingly popular and important as technology makes such polling easier.

Generally, polling, as a form of direct democracy is effective for issues which are relatively simple. and about which the silent majority have an opinion that is under-represented. For more complex issues, such direct democracy is criticized as populist and irresponsible.

To address this issue, Professor James S. Fishkin, Director of Stanford University’s Center for Deliberative Democracy, has developed a method called deliberative polling. Deliberative polling combines deliberation in small group discussions with scientific random sampling to increase the quality and depth of the understanding of the participants, while maintaining a sampling that reflects the actual distribution of opinion in the population, rather than the distribution of political power. Deliberative polling has been used successfully to poll people about relatively complex issues such as tax policies \cite{11}.

\section*{Emergence}

Emergence is a term relevant to the study of complex systems. Emergence is what you have when the relatively simple interactions of relatively simple parts of a system yield complex results over time. Emergent behaviors are behaviors that are not directed by systems of command and control, but emerge from subtle, complex interactions. Common examples are flocks of ducks or birds that act in concert but with no specific leader, or colonies of ants that establish routes for collecting food based on group experience reinforced by pheromones.

In the book Emergence \cite{12}, Steven Johnson writes about harvester ant colonies, which exhibit an amazing ability to solve difficult problems, including geometry. The following exchange is from an interview with Deborah Gordon who studies ants.

\begin{quotation}
She says, ``Look at what actually happened here: they've built the cemetery at exactly the point that's furthest away from the colony. And the midden is even more interesting: they've put it at precisely the point that maximizes its distance from both the colony and the cemetery. It's like there's a rule they're following: put the dead ants as far away as possible, and put the midden as far away as possible without putting it near the dead ants.''
\end{quotation}

Johnson explains that there is no single ant in charge. The ants' solving of such problems is emergent behavior resulting from simple rules and diverse interactions with immediate surroundings and neighbors.

The complex human fetus develops from simple cells through this same principle: following a simple set of rules encoded in DNA. When the first cell divides into two, one half becomes the head side and the other the tail. The next time it divides, the quarters determine whether they are to be the head or the tail, and they become the head of the head, or the tail of the head, and so on. This division and specialization continues until in very short order the cells have created a complex human body. The liver cells know to turn into liver cells by sensing that their neighbors are also liver cells and “reading” the DNA. There is no omniscient control, just a growing number of independent cells following rules and communicating with and sensing the state of their neighbors \cite{13}.

In The Death and Life of Great American Cities, Jane Jacobs argues that urban planning in America tends to fail when top-down plans to change the nature of neighborhoods are implemented. Most large projects designed to increase the quality of ghetto areas by building large apartment projects have not succeeded in their aim. Conversely, neighborhoods that have thrived have done so through a kind of emergence. She argues that the interaction between people on the sidewalks and streets creates a street culture and intelligence more suitable than central control for managing neighborhoods in cities, and that instead of bulldozing problems, city planners should study neighborhoods that work and try to mimic the conditions that produce the positive emergent behavior \cite{14}.

Can citizens self-organize to deliberate on, and to address, complex issues democratically, without any one citizen required to know and comprehend the whole? This is the essence of emergence, the way that ant colonies can ``think'' and cellular DNA can evolve complex human bodies. If information technology could provide tools for citizens in a democracy to participate and interact in a way that facilitates self-organization and emergent understanding, we can evolve a form of emergent democracy that would resolve complexity and scalability issues associated with democratic governance.

In complex systems the role of the leader is not about determining direction and controlling followers. The leader maintains integrity, mediates the will of the many, influencing and communicating with peers and with other leaders \cite{15}. The leader becomes more of facilitator (or hub), and custodian of the process, than a power figure. She is the catalyst or manager of critical debate, or the representative of a group engaged in critical debate \cite{16}. The leader is often the messenger delivering the consensus of a community to another layer or group. As leadership becomes necessary to manage the development of an opinion or idea about a complex issue, information technology can enable quick and ad hoc leader selection and representation of consensus opinion in a larger debate.

\section*{Weblogs and emergence}

In \emph{Emergence}, Steven Johnson writes:

\begin{quotation}
The technologies behind the Internet--everything from micro-processors in each Web server to the open-ended protocols that govern the data itself--have been brilliantly engineered to handle dramatic increases in scale, but they are indifferent, if not down-right hostile, to the task of creating higher-level order. There is, of course, a neurological equivalent of the Web's ratio of growth to order, but it's nothing you'd want to emulate. It's called a brain tumor.
\end{quotation}

\emph{Emergence} was written in 2001. A change has taken place on the Internet since 2000. Weblogs, which we have defined as personal web sites with serial content posted in reverse chronological order, have begun to grow in number and influence. Weblogs exhibit a growing ability to manage a variety of tasks, and emergent behavior is evident because of changes in the way weblogs are managed.

Johnson's explanation for the inability of web pages to self-organize is,

\begin{quotation}
Self-organizing systems use feedback to bootstrap themselves into a more orderly structure. And given the Web's feedback-intolerant, one-way linking, there's no way for the network to learn as it grows, which is why it's now so dependent on search engines to rein in its natural chaos.
\end{quotation}

He also describes how, in the example of the ants, the many simple, local, random interactions of the ants helped them exhibit emergent behavior.

Weblogs are different from traditional web pages in several ways. Weblogs involve the use of content management tools, which make it much easier to add entries, with a resulting increase in the number and frequency of items posted. The posts are generally small items with a variety of information types - e.g. text, photographs, audio, and video referred to as micro-content \cite{17}. Weblog culture encourages bloggers (people who run weblogs) to comment on entries in other weblogs and link to the source. Some systems have a protocol that supports interactive linking: i.e. when a blogger posts an item with a link to another weblog, a link to his new item is created on that weblog. In addition to HTML content, weblogs often generate XML \cite{18}. files based on a standard protocol for syndication called RSS \cite{19}, which allows computers to receive updates to weblogs through special clients aggregating syndicated content - such as Feedreader \cite{20}, for Windows and NetNewsWire \cite{21} for the Macintosh. These news aggregators constantly scan the users' favorite weblogs for new posts.

When new entries are posted to a weblog, a notification may also be sent to services such as weblogs.com \cite{22}, which keep track of weblog updates in near real-time. This information is also used by a variety of new services to generate meta-information about weblogs. These new information sites include Blogdex \cite{23}, which scans weblogs for quoted articles and ranks them according to the number of weblog references, and Technorati \cite{24}, which ranks weblogs by tracking inbound and outbound links to specific weblogs and/or weblog posts.

Technorati's results in particular look like diagrams of small-world networks \cite{25}, Weblog links are governed by much the same rules. They represent a scale-free network of weblogs where friends generally link to friends, but some weblogs serve as hubs with many more connections, including links to whole other clusters of weblogs, and to other content within the Internet. (It would be interesting to see how the pattern of weblog links looks relative to linking patterns in the web overall. Are weblogs an organizing structure of the web, or merely another cluster within the web?)

In addition to linking articles between weblogs, bloggers link to each other via blogrolls, marginal lists of personal favorite weblogs. Services such as blogrolling.com \cite{26} help bloggers manage their blogrolls and see who is blogrolling them. Services such as blogstreet \cite{27} provide a method of viewing the ``neighborhood'' of a blogger by following and analyzing blogroll links.

In this way, the structure of weblogs addresses the problem that Johnson raised when he suggested that the Web is not self-organizing. Through the feedback and two-way linking we have described, weblogs show emergent self-organization.

\section*{The Power Law}

With the appearance of the World Wide Web, proponents hoped that the low barriers to entry (inexpensive web hosting, ease of setting up a web page) would dramatically increase the number of people publishing their thoughts, and that this would lead to a diverse and decentralized system. What happened instead was that portals and search engines captured much of the traffic and an attention economy \cite{28} formed as attention became a scarce resource for which various commercial entities competed. Users focused on portals first to help them find what they were looking for. Then they went to the large ecommerce and news sites that appeared during the Internet boom. These sites provided a sense of order, a variety of products, and high quality information. A minority of web surfers landed on smaller, less prominent sites. This attention economy created a value in site traffic, which was purchased from more popular sites in the form of paid advertisements and sponsored links. This is still the primary income model for search engines and portal sites today.

In a widely distributed and linked paper, Clay Shirky argues that weblogs are exhibiting a sort of order now because the community is still small. As the community increases in size, he contends, this order will fragment, as it did for online communities in the past, such as Usenet news groups, mailing lists and bulletin boards. In his paper, ``Power Laws, Weblogs, and Inequality,'' \cite{29} Shirky shows that an analysis of inbound links for weblogs shows a standard power law distribution. The power law distribution is a distribution where the value of any unit is 1/n of its ranking. The second place weblog has 1/2 of the inbound links of the top ranking weblog, the third place weblog has 1/3 of the inbound links and so on.

This power law distribution can be counterintuitive. Shirky argues that the top-ranking weblogs will eventually become mass media, while the weblogs at the bottom of the curve will have difficulty gaining any attention. As a result, these weblogs will appear as nothing more than local conversations with friends. He suggests that it will be increasingly difficult to displace the high-ranking sites, and his power law distribution data for weblogs supports his claims.

Shirky’s analysis may be missing important factors, however. Weblogs form a scale-free network where some nodes are hubs, i.e. more heavily linked than others, and this does suggest a power law distribution. However there may be dynamism that the power law doesn’t capture. Subnetworks of weblogs may become linked, for instance, as during the Iraqi war, when warbloggers (a subset or subnetwork of bloggers supporting the war) debated antiwar bloggers, thereby forming links between the two networks. This has resonance with the concept of emergent communities of interest espoused by Valdis Krebs, which demonstrates how subnetworks may be linked through affinity points \cite{30}.

\section*{Mayfield's Ecosystem}

Ross Mayfield, CEO of the social software company SocialText, proposed an alternative view of the political economy of weblogs. Mayfield points out that not all links have equal value. He explains that there are three different types of networks developing among weblogs: creative, social, and political networks.

A creative network is a flat network of a production-oriented group of close associates with deep trust and dense inter-linking. It is said that 12 people is the optimum number for holding a dinner conversation or a tight team \cite{31}.

A social network is the traditional weblog form. The Law of 150 \cite{32} is a theory that people can maintain an average of 150 personal relationships. The Law of 150 is a bell-shaped distribution where some weblogs receive more attention than others, but the distribution fairly represents the quality of the weblogs.

A political network follows Shirky's power law and is similar to a representative democracy where weblogs receive links from thousands of other weblogs. Each link may be thought of as a vote. The weblogs at the top of this power curve have a great deal of influence.

\section*{Strength of Weak Ties}

In ``The Strength of Weak Ties,'' Mark Granovetter \cite{33} describes the value of weak ties in networks. Strong ties are your family, friends and other people you have strong bonds to. Weak ties are relationships that transcend local relationship boundaries both socially and geographically. A study by Granovetter demonstrates that people are more likely to find employment through their weak ties than their strong ties.

It is the ability to operate in all three of Mayfield's clusters, and to transcend boundaries between them that make weblogs so potentially powerful. A single weblog and even a single entry in a weblog can have an operational purpose, a social purpose, and an impact on the political network. Recall that emergence seems predicated on many mechanisms of communication between elements. For instance, when I blog something about Emergent Democracy, I may be speaking creatively to the small group of researchers working on this paper; socially to a larger group of friends who are thinking along with me and trying to get a handle on the concept; and on a political level to readers I don’t know, but who I’m hoping to influence with my talk about a new kind of politics.

Many bloggers create their weblogs in order to communicate with their strong-tie peers, linking to and communicating within this small group at the creative level. At some point, someone in the peer group will discover some piece of information or point of view which resonates with the next, social level. Then a larger number of social acquaintances will pick up those entries that they believe may be interesting to others in their individual social networks. In this way, a small group focusing on a very specific topic can trigger a weak-tie connection carrying useful information to the next level. If this information resonates with even more bloggers, the attention given the source will increase rapidly. The individual or group who created the original comment or post will also continue to participate in the conversation, since they can be aware, through Technorati or blogdex, of all of the links to the original piece of information as they propagate.

Weblogs create a positive feedback system, and with tools for analysis like Technorati, we can identify the importance of information at the political level by tracking its movement across the weak ties between networks and network levels.

Noise in the system is suppressed, and signal amplified. Peers read the operational chatter at Mayfield's creative network layer. At the social network layer, bloggers scan the weblogs of their 150 acquaintances and pass the information they deem significant up to the political networks. The political networks have a variety of local maxima which represent yet another layer. Because of the six degrees phenomenon, it requires very few links before a globally significant item has made it to the top of the power curve. This allows a great deal of specialization and diversity to exist at the creative layer without causing disruptive noise at the political layer. The classic case, already mentioned above, was the significant chatter at the creative level when Trent Lott praised Strom Thurmond's 1948 segregationist campaign for the presidency, though conventional journalists had ignored the comment. The story escalated to influential bloggers and there was a real impact at the political level, leading to Lott’s resignation.

\section*{The brain and excitatory networks}

For a couple years now, software engineer Peter Kaminski of SocialText \cite{34} has been working on the hypothesis that the process that governs the way we think described by neurobiologist author William Calvin \cite{35} as the ``emergent properties of recurrent excitatory networks in the superficial layers of cerebral cortex,'' scales up in a similar fashion to the way people work together in groups, and groups of groups -- and ultimately, up to direct democracy \cite{36}.

Calvin notes that the cerebral cortex is made up of columns of neurons, which are tightly interlinked and analogous to the creative network. These columns resonate with certain types of input. When they are excited, they excite neighboring columns. If the neighboring columns resonate with the same pattern, they also excite their neighbors. In this way, the surface of the cerebral cortex acts as a voting space, with each column of neurons, when excited by any of a variety of different patterns (thoughts), selectively resonating and then exciting their neighbors. When a significant number of the columns resonate with the same pattern, the thought becomes an understanding. Sensory organs provide the inputs to the various columns and the brain as a whole drives output to other organs based on the understanding.

Calvin's model of human thought process suggests that the brain uses emergence, the strength of weak ties, and a neighbor excitation model for resolving thoughts. The structure of the brain is similar to Mayfield's system. One of the keys is that the columns only excite their neighbors. This self-limiting factor is also one of the factors that Johnson describes in creating the emergent behavior of ants. The influence of weblogs is similarly constrained by the ability of individuals to read only a limited number of weblog entries per day and the tendency to focus, not on the weblogs with a high political ranking, but on the creative and social weblogs of interest. This dampening feedback is essential in maintaining the volume of interaction in the important zone of maximum emergence between completely random noise and completely useless order.

\section*{Trust}

Another important aspect of understanding the relationship between the components of the network and the nature of emergent behavior in human networks is the issue of trust.

Francis Fukuyama, in his book Trust \cite{37}, says that it was the nations that managed to create a layer of trust larger than the family unit and smaller than the nation that were able to build large and scalable organizations. In pre-industrial Germany, it was the guilds, in early Japan it was the \emph{iyemoto} (feudal families which allowed new members), and in the US, it was a variety of religious groups.

Behavioral psychologist Toshio Yamagishi \cite{38} distinguishes between assurance and trust \cite{39}. Yamagishi argues that, in a closed society, people do not base their social expectations on trust. Rather, behavioral standards derive from the inability of the individual to escape from the community, and the fear of punishment. Conversely in open communities where people are free to come and go, trust and trustworthiness are essential to creating collaborative organizations. Yamagishi provides data showing that closed societies such as Japan have a lower percentage of people who trust others than open societies, such as the United States, where trust between individuals is necessary.

Yamagishi conducted an experiment using a market simulation where participants were classified as buyers or sellers. They bought and sold items within their groups. The sellers could lie about the quality of the items that they were selling. In the closed market scenario where sellers’ behaviors were associated with their identities, the quality of the transactions was naturally high. In a completely anonymous system, the quality was low. When participants were allowed to change their identities and only negative reputation was tracked, the quality started high but diminished over time. When the participants were allowed to change their identities and only positive reputation was tracked, the quality started low but increased over time and approached the quality of transactions in the closed network \cite{40}.

As networks become more open and complex, the closed networks which rely on the ability to punish members and the ability to exclude unknown participants becomes extremely limiting. The dynamic open networks, which rely on the ability of members to trust each other and identify trustworthiness through reputation management, are scalable and flexible. Links between weblogs, the ability to view the histories of individuals through their weblogs and the persistence of the entries enhances greatly the ability to track positive reputation. Trust and reputation build as the creative, social and political networks harbor mutual respect recognized and illustrated through linking and reciprocal linking, particularly in blogrolling behavior and secondarily in linking and quoting. Another factor in maintaining a high level of trust is to create an ethic of trustworthiness. Trustworthiness comes from self-esteem, which involves motivation through trusting oneself rather than motivation through fear and shame \cite{41}.

\section*{The toolmakers}

After the Internet bubble a great number of talented programmers and architects were no longer focused on building components for large projects, which were often doomed by the basic top-down nature of hastily built business plans concerned more with investor appeal than anything else. These talented programmers and architects are now more focused on smaller projects to build tools and design architectures for themselves, instead of creating innovative technologies for imagined customers in imagined markets for investors imagining valuations and exits. These toolmakers are using tools they have created to communicate, discuss, and design new infrastructures. They are sharing information, setting standards, and collaborating on compatibility. The community of toolmakers for weblogs and associated technology is vibrant, similar to the Internet Engineering Task Force (IETF) during the early days of the Internet, when independent programmers were first allowed to write networking software and enter the domain previously controlled by large hardware companies and telecommunications firms.

The weblog developer community, which initially developed tools for personal use, now has significant impact and influence on mass media, politics, and classic business networking. This inspires hope that we will discover how to scale the weblog network in a way that will allow bloggers to play an increasingly important role in society.

\section*{Where are we today?}

There are several million weblogs on the Internet. However, the tools are still difficult for many people to use, and weblogs are still an obscure phenomenon, especially for those who spend little or no time online. However more weblogs appear every day, and what started as an American phenomenon is rapidly beginning to appear in other countries.

One aspect of weblogs that has increased their value over traditional web pages is the frequency and immediacy of discussion. Recently, a group of bloggers, including myself, have started to organize ``Happenings'' \cite{42}, which involve a live voice conference, a chat room for parallel conversation and for moderating the voice conference, and a Wiki, a tool that allows any number of people to easily create and edit plain-text web pages ) in order to provide a space for collaboration. Weblogs by their nature can be updated as fast as email, but chat and voice provide faster and more personal levels of communication as the discussion of an issue expands and escalates from the creative to the political level.

With the increase in wireless mobile devices like cameras and phones, mobile weblogging, or ``moblogging'' \cite{43} (posting photos and text from mobile phones and other mobile devices )is gaining popularity. As location information becomes available for the mobile devices, moblogging will be a way to annotate the real world, allowing people to leave information in locations or search for information about specific locations. Although moblogging has privacy issues, its ability to contribute to Steve Mann's vision of sousveillance is significant. Sousveillance, French for ``undersight,'' is the opposite of surveillance. Examples of sousveillance include citizens keeping watch on their government and police forces, student evaluations of professors, shoppers keeping tabs on shopkeepers \cite{44}.

All of these new developments are components that are being tied together with open standards and a community of active architects and programmers. A dialog, tools, and a process to manage this dialog is emerging.

This paper was written using this process. A variety of people were engaged in conversations on weblogs about democracy, weblog tools, critical debate, the war in Iraq, privacy and other issues discussed in this paper. As these ideas were linked across weblogs, a group of people resonated with the idea of emergent democracy. I asked people to join me in a telephone call and we had an initial voice conference call of about twelve people where we identified some of the primary issues. Ross Mayfield called it a ``happening.''

We scheduled another call, which included 20 people, and many of the people from the first call provided tools to support the happening, including a Wiki; a trackback weblog \cite{45}, which tracked entries in different weblogs about emergent democracy; a chat; and a teleconference that was open for anyone to call and join the discussion. The second happening moved the discussion to the next level of order, and, as a result, I was able to organize some of the thoughts into the first draft of this paper.

I posted the draft of this paper on my weblog \cite{46} and received a great number of comments and corrections, which sparked another email dialog about related topics. Much of this feedback has been integrated into this version of the paper, which is my version of a community dialog on the Internet, and could not have been written without this community’s input and access to the social tools described above.

\section*{Conclusion}

We have explored the concepts of democracy and emergence, how they are related, and how practical applications of the two concepts are supported by social technologies. The authors feel that the emergent democracy provides an effective next step toward a more participatory form of government that leverages the substantial advances in communications technology that we’ve seen over the last century. Traditional forms of representative democracy can barely manage the scale, complexity and speed of the issues in the world today. Representatives of sovereign nations negotiating with each other in global dialog are limited in their ability to solve global issues. The monolithic media and its increasingly simplistic representation of the world cannot provide the competition of ideas necessary to reach informed, viable consensus. The community of developers building social software and other tools for communication should be encouraged to consider their potential positive effect on the democratic process as well as the risk of enabling emergent terrorism, mob rule and a surveillance society.

We must protect the availability of these tools to the public by protecting the electronic commons. We must open communications spectrum and make it available to all people, while resisting increased control of intellectual property, and the implementation of architectures that are not inclusive and open. We must work to provide access to the Internet for more people by making tools and infrastructure cheaper and easier to use, and by providing education and training.

Finally, we must explore the way this new form of democratic dialog translates into action and how it interacts with the existing political system. We can bootstrap emergent democracy using existing and evolving tools and create concrete examples of emergent democracy, such as intentional blog communities, ad hoc advocacy coalitions, and activist networks. These examples will create the foundation for understanding how emergent democracy can be integrated into society generally.

\section*{Acknowledgements}

Special thanks to all of the people who participated in the happening, sent me suggests and commented on my weblog regarding this paper. These people include: Clay Shirky, Ross Mayfield, Pete Kaminski, Gen Kanai, Liz Lawley, Sébastien Paquet, Flemming Funch, Adina Levin, Edward Vielmetti, Greg Elin, Stuart Henshall, Jon Lebkowsky, Florian Brody, Mitch Ratcliffe, Kevin Marks, George Por, Dan Gillmor, Allan Karl, Rich Persaud , George Dafermos, Steve Mann, Karl-Friedrich Lenz , Toph, Chris Case and Howard Rheingold.

\bibliography{joi}

\begin{thebibliography}{10}
\bibitem{18}
Extensible markup language ({XML}) is a simple, very flexible text format
  derived from sgml ( iso 8879 ). originally designed to meet the challenges of
  large-scale electronic publishing, {XML} is also playing an increasingly
  important role in the exchange of a wide variety of data on the web and
  elsewhere.
\bibitem{19}
{RSS} is a web content syndication format. its name is an acronym for {Really}
  {Simple} {Syndication}. {RSS} is dialect of {XML}. all rss files must conform
  to the {XML} 1.0 specification, as published on the world wide web consortium
  ({W3C}) website. at the top level, a rss document is a <rss> element, with a
  mandatory attribute called version that specifies the version of rss that the
  document conforms to.
\end{thebibliography}

\end{document}